\documentclass[ prl,  twocolumn,  superscriptaddress]{revtex4-1} %twocolumn/preprint,  groupedaddress/superscriptaddress, longbibliography, linenumbers
\usepackage{graphicx}
\usepackage{epstopdf}			% Use eps figures
\usepackage{color}
\usepackage{amsmath} % split environment
\usepackage{amsfonts}
\usepackage{mathtools} % multlined environment
\usepackage{mwe,tikz}\usepackage[percent]{overpic}
\usepackage{tabularx}
\usepackage{diagbox}
\usepackage{hyperref}

\newcommand{\kT}{k_\mathrm{B}T}

\begin{document}
%\linenumbers

\title{Predicting Rubisco:Linker Condensation from Titration  in the Dilute Phase }
%\title{Predicting Rubisco:Linker Condensation from Single-Rubisco Assemblies } 
\date{\today}
\author{Alex Payne-Dwyer}
\thanks{These authors contributed equally}
\affiliation{School of Physics, Engineering and Technology, University of York, York, YO10 5DD, United Kingdom}
\author{Gaurav Kumar}
\thanks{These authors contributed equally}
\affiliation{Department of Biology, University of York, York, YO10 5DD, United Kingdom}
\affiliation{Centre for Novel Agricultural Products (CNAP), Department of Biology, University of York, York, YO10 5DD, United Kingdom}
\author{James Barrett}
\thanks{These authors contributed equally}
\affiliation{Department of Biology, University of York, York, YO10 5DD, United Kingdom}
\affiliation{Centre for Novel Agricultural Products (CNAP), Department of Biology, University of York, York, YO10 5DD, United Kingdom}
\author{Laura K. Gherman}
\affiliation{Department of Biology, University of York, York, YO10 5DD, United Kingdom}
\affiliation{York Structural Biology Laboratory, The University of York; York, YO10 5DD, UK}
\author{Michael Hodgkinson}
\affiliation{Department of Biology, University of York, York, YO10 5DD, United Kingdom}
\author{Michael Plevin}
\affiliation{Department of Biology, University of York, York, YO10 5DD, United Kingdom}
\affiliation{York Structural Biology Laboratory, The University of York; York, YO10 5DD, UK}
\author{Luke Mackinder}
\affiliation{Department of Biology, University of York, York, YO10 5DD, United Kingdom}
\affiliation{Centre for Novel Agricultural Products (CNAP), Department of Biology, University of York, York, YO10 5DD, United Kingdom}
\author{Mark C. Leake}
\affiliation{School of Physics, Engineering and Technology, University of York, York, YO10 5DD, United Kingdom}
\affiliation{Department of Biology, University of York, York, YO10 5DD, United Kingdom}
\author{Charley Schaefer}
\email{charley.schaefer@york.ac.uk}
\affiliation{School of Physics, Engineering and Technology, University of York, York, YO10 5DD, United Kingdom}

%https://www.overleaf.com/project/6421a56e9ac5b5f42f98a820

\begin{abstract}
{ 
The condensation of Rubisco holoenzymes and linker proteins into `pyrenoids', a crucial supercharger of photosynthesis in algae, is  qualitatively understood in terms of `sticker-and-spacer' theory.
We derive semi-analytical partition sums for small Rubisco:linker aggregates, which enable the calculation of both dilute-phase titration curves and dimerisation diagrams. 
By fitting the titration curves  to Surface Plasmon Resonance and Single-Molecule Fluorescence Microscopy data, we extract the molecular properties needed to predict dimerisation diagrams.
We use these to estimate typical concentrations for condensation, and successfully compare these to microscopy  observations. 
}\end{abstract}

\maketitle
%\tableofcontents
%

%\section{Introduction}
 %\cite{SchaeferInterface}

Biopolymer networks are ubiquitous in nature as biomaterials such as silk \cite{Schaefer20, Schaefer22} and artificial hydrogels \cite{Hughes21}, and fulfil vital physiological roles intra- and extracellularly \cite{SchaeferInterface, Fosado23}, in particular as natural  \cite{Shin17, choi_physical_2020, JinX20, Connor22, Qian22} and  artificial \cite{Lasker22} (multicomponent \cite{Harmon17, Zhang21, Qian22}) biomolecular condensates.
Several  emergent  properties of self-assembly can be reproduced by  models through  interactions via linker molecules across a biopolymer network, often proteins with high levels of intrinsic disorder, which comprise ‘stickers’ interspersed by ‘spacers’ \cite{choi_physical_2020}.
The most crucial molecular properties, the sticker binding affinity and the extensibility of the spacers, are both typically unknown. Consequently,  the need of extensive simulation assays \cite{GrandPre23} imposes a practical challenge to falsify or advance the theory.
In this Letter, we remedy this by deriving semi-analytical solutions that enable the full parametrisation in the dilute phase, as well as the computationally efficient calculation of dimerisation diagrams.

\begin{figure}[ht!]
\centering
\includegraphics*[width=8cm]{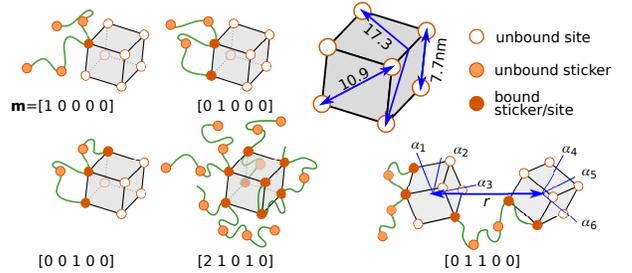}%\vspace{-0.2cm}
\caption{Rubisco is parametrised using a cube with a space diagonal of  $13.4$ nm  and whose corners represent binding sites \cite{Note1}. The permutations of linker binding are described using integer partitions' $\mathbf{m}$, and the conformations using rotations $\boldsymbol{\alpha}=[\alpha_1,\dots,\alpha_6]$ and centre-to-centre distance $r$.}
\label{fig:ModelSchematics}
\end{figure}

As a model system we focus on  the `pyrenoid', which is a phase-separated organelle found in the photosynthetic chloroplast of eukaryotic algae and some basal land plants \cite{Mackinder16, He20, Barrett21}. 
Across species, the supercharging of photosynthesis relies on the  crosslinking of the principal CO$_{2}$-fixing holoenzyme Rubisco (schematically represented by the cube in Fig.~\ref{fig:ModelSchematics}) by multivalent linkers,  whose binding motifs  may bind to $8$ specific sites on Rubisco \cite{Meyer20, He20}.
Binding is reversible, as evidenced by the liquid-like properties of the pyrenoid that were found \emph{in vivo} and \emph{in vitro} by rapid internal mixing, fusion and fission  \cite{Freeman17, Wunder18}.
The most widely studied species is the model green alga \emph{C. reinhardtii}, whose linker protein Essential Pyrenoid Component 1 (EPYC1) has $5$ `Rubisco binding motif' stickers \cite{Meyer20, He20} that facilitate multiplicit binding \cite{He23}.

In the following, we will derive semi-analytical partition sums for Rubisco monomers and dimers.
We then use the monomeric partition sum to calculate titration curves that we fit to Surface Plasmon Resonance (SPR) and single-molecule fluorescence  microscopy (Slimfield) data.
This yields quantitative values for the sticker binding energy $\varepsilon$ and the Kuhn length of the spacers, $l_\mathrm{K}$.
We use these to calculate dimerisation diagrams using the dimeric partition sum, and compare the theoretical predictions to microscopy observations of droplet formation.

 \vspace{-0.2cm}

\paragraph*{Equilibrium Self-Assembly Theory. --} 
 We model Rubisco as a cubic patchy particle \cite{Zhang04} with at each corner a site  to which stickers may bind with an  energy $\varepsilon$. If stickers $i$ and $j>i+1$ bind to two sites at a distance $z$  (see Fig.~\ref{fig:ModelSchematics}), where stickers $i<k<j$ are open, a strand of  $n_{ij}=\sum_{k=i}^{j-1}n_k$ monomers is stretched, with $n_k$  the number of amino acids between stickers $k-1$ and $k$ (we fix $n_k=50$ in this Letter). We describe the entropic penalty of stretching the $n\equiv n_{ij}$ strand using the freely-jointed chain model \cite{Schieber03} \footnote{Supplementary Materials will be made available following publication.},
\begin{align}
  \frac{G(z;n)}{\kT} &= \frac{3}{2}\ln\left(\frac{2\pi \beta n_\mathrm{K} }{3} \right) \notag\\
  &+ 
       n_\mathrm{K}\left[ \frac{1}{2}\left(\frac{z}{bn}\right)^2 -  
  \ln\left(
    1-\left(\frac{z}{bn}\right)^2
  \right)\right],  \label{eq:StretchPotential}
\end{align}
with $b=0.36$ nm the step length of an amino acid; $\kT$ the thermal energy and $n_\mathrm{K}=nb/l_\mathrm{K}$ the number of Kuhn segments, with $l_\mathrm{K}$ the Kuhn length (typically $0.36$ to $1.5$ nm for polypeptides \cite{MullerSpath10,Cheng10,Zahn16,Schaefer20}).
We assume that the non-universal constant $\beta$\cite{Schieber03} equals unity, which ensures a positive entropic penalty  for any value for $z$.% \footnotemark[\ref{note1}]

We will consider all binding permutations to calculate the partition sum of single-Rubisco complexes,
%\begin{equation}
$Z^{(1)}=\sum_{M=0}^{N}\mathrm{e}^{M \mu/\kT} Z_M^{(1)} $ % $,\label{eq:Z}
% \end{equation} 
where $N=8$ is the number of sites per Rubisco and $\mu$ is the  chemical potential of the linkers.
$Z_{M}^{(1)}$ is the partition sum of binding $M$ linkers, and enables the calculation of titration curves through the $\mu-$dependent mean number of bound linkers,
\begin{equation}
  \langle M\rangle = \frac{\sum_{M=1}^{N} M \mathrm{e}^{M \mu/\kT} Z_{M}^{(1)}}{\sum_{M=0}^{N} \mathrm{e}^{M \mu/\kT}Z_{M}^{(1)}}.\label{eq:Mmean}
\end{equation}
We found \cite{Note1} that $Z_{M}^{(1)}$ can be written as
\begin{equation}
  Z_{M}^{(1)} = Z_{\mathrm{trans}}^{(1)}Z_\mathrm{rot}^{(1)}\sum_{B=M}^{\mathrm{min}\left\{N,MS\right\}}\mathrm{e}^{-B\varepsilon/\kT}Z_{{M,B}}^{(1)},\label{eq:ZBM1}
\end{equation}
with $Z_{\mathrm{trans}}^{(1)}$ and $Z_\mathrm{rot}^{(1)}$  the translation and rotational partition sums, and with
\begin{equation} 
Z_{{M,B}}^{(1)}
=\sum_{\mathbf{m}\in \mathcal{P}_{M,B}} \psi_\mathbf{m}\Omega_\mathbf{m}^0\langle \exp(-G_\mathrm{elas}/\kT)\rangle_\mathbf{m},
\label{eq:ZMB}
\end{equation}
 for  $B\in [M,\,\min\{N,MS\}]$ bound stickers, with 
$\mathcal{P}_{M,B}$ the set of  `\emph{integer partitions} ' $\mathbf{m}\equiv [m_1,\, m_2,\dots,\,m_{S}]$ whose elements $m_b$ count the number of molecules that are bound using $b$ stickers. The set is found using an algorithm \cite{Stoj98, Note1}.
For each integer partition (for $S=5$ there are $64$ of them), the upper limit for the number binding configurations is, 
\begin{equation}
  \Omega_\mathbf{m}^0=
   \frac{N!}{(N-B)!} \prod_{b=1}^{S}\frac{1}{m_b!}\left(\frac{S!}{(S-b)!b!}\right)^{m_b},
\end{equation}
and which is damped by the factor $\psi_\mathbf{m}\leq 1$ to correct for physically inaccessible states due to the overstretching of spacer strands.
The final term in Eq.~(\ref{eq:ZMB}), $\langle \exp(-G_\mathrm{elas}/\kT)\rangle_\mathbf{m}$, is the ensemble averaged Boltzmann factor due to the spacer entropy in Eq.~(\ref{eq:StretchPotential}).
 $\psi_\mathbf{m}$ and $\langle \exp(-G_\mathrm{elas}/\kT)\rangle_\mathbf{m}$ are obtained through numerical sampling \cite{Note1}.

\begin{figure}[ht!]
\centering
\includegraphics*[width=9.0cm]{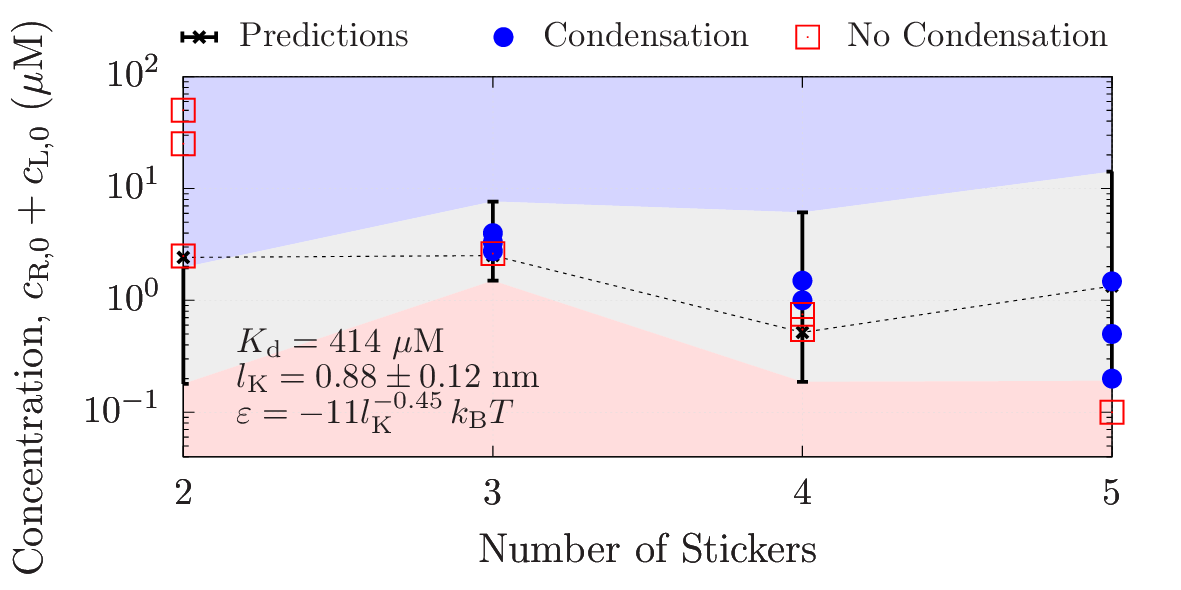}
\caption{
    Characteristic concentration  for self-assembly against the number of stickers per linker molecule. The predictions are based on the best fit to the single-molecule data (dashed line) and the propagated uncertainty (grey shaded area). The microscopy observations are summarised by the open and closed symbols.
  }
  \label{fig:StickerDependence}
\end{figure}

In analogy with the monomeric partition sum, for the dimer we have $Z^{(2)} = \sum_{M=1}^{2N-1}\exp(M\mu/\kT)Z_M^{(2)}$, and which allows for the calculation of the dimerisation constant $K_\mathrm{a} = { \lambda^3 Z^{(2)}}/{\left(Z^{(1)}\right)^2}$, with $\lambda=\sqrt{h^2/2\pi m\kT}$ the thermal wavelength, $h$ Planck's constant and $m$ the mass of Rubisco. This gives for the fraction of dimers $f = 1 - ({4K_\mathrm{a}c_\mathrm{R,0}})^{-1}\left(\sqrt{1+8 K_\mathrm{a}c_\mathrm{R,0}}-1\right)$, with $c_\mathrm{R,0}$ the Rubisco concentration \cite{Note1}. We will use the dimerisation reaction as a proxy for condensation by assuming droplet formation is not nucleated. In this spirit we approximate the `spinodal branch' by the condition where half of the material is dimerised, $f=1/2$, and approximate the critical concentration (which we compare to experimental approximates in Fig.~\ref{fig:StickerDependence}) by the `lower dimerisation concentration' of Rubisco and linker for which this holds,
\begin{equation}
  c = \min\left\{c_\mathrm{R,0}+c_\mathrm{L,0}\, \middle| \,f(c_\mathrm{R,0},c_\mathrm{L,0})=1/2\right\}.\label{eq:DimerisationConcentration}
\end{equation}

To calculate $Z^{(2)}$, we take into account that the intersite distances depend on the six axes of rotation, $\boldsymbol{\alpha}$, of both Rubisco monomers, two axes of rotation of the dimer, and  the vibrational modes described by the centre-to-centre distance $r$, see Fig.~\ref{fig:ModelSchematics}. 
We therefore write
%\begin{widetext}
\begin{align}
  Z_M^{(2)} =&  \frac{2Z_{\mathrm{trans}}^{(1)}}{\lambda}\int \mathrm{d}r Z_\mathrm{rot}^{(2)}(r)  \,  
  \left(\frac{Z_\mathrm{rot}^{(1)}}{4\pi^3}\right)^{2}\\\notag
  &  \times \int \mathrm{d}\boldsymbol{\alpha}
    \sum_{B=M+1}^{\min\{2N-1,MS\}} 
    \mathrm{e}^{-\beta \varepsilon B} \\\notag
    & \times \left( Z^{(2)}_{M,B}(r,\boldsymbol{\alpha})-\sum_{m,b}Z^{(1)}_{m,b}Z^{(1)}_{M-m,B-b}\right)\Theta(r,\boldsymbol{\alpha}),
\end{align}
%\end{widetext}
where $\Theta(r,\boldsymbol{\alpha})$ is zero if the Rubiscos intersect and unity otherwise. We used $Z_{\mathrm{trans}}^{(2)}=2^{3/2}Z_{\mathrm{trans}}^{(1)}$ and  $\lambda_\mathrm{vib}=2^{1/2}\lambda$ \cite{Klein18}, and 
$Z_\mathrm{rot}^{(2)}= 8\pi^2I\kT/2h^2= \pi r^2/\lambda^2$ is the dimeric rotational partition sum   with $I=mr^2/2$ the moment of inertia \cite{Zangi22}.
The angular integrals double count the orientations already captured by $Z_\mathrm{rot}^{(1)}$, and are corrected for by the factor $4\pi^3$ (we refer to Ref.~\citenum{Cates15} for an excellent discussion on the `entanglement' between rotations and permutations).
For each orientation the permutations of binding linkers is described by $Z^{(2)}_{M,B}(r,\boldsymbol{\alpha})$, and requires the subtraction of all monomeric states in the summation $\sum_{m,b}$. 
It is infeasible to calculate $Z^{(2)}_{M,B}$ through sampling as in Eq.~(\ref{eq:ZMB}), because $\psi_\mathbf{m}\ll 1$ due to the relatively large intersite distances.
However, because for weak binding ($\varepsilon\rightarrow \infty$)   any molecule  binds only using a single sticker, $\lim_{\varepsilon\rightarrow \infty}Z^{(2)}_{M,B}=S^M(2N)!/[(2N-M)!M!]$, and in general we can calculate $Z^{(2)}_{M,B}$ using thermodynamic integration with $\varepsilon$ as the integration variable. This gives \cite{Note1},
\begin{equation}
  Z_{M}^{(2)} = \frac{\pi}{\lambda^3}
   \,    \mathrm{e}^{-M\beta\varepsilon}\int \mathrm{d}r r^2 f_\mathrm{ex}(r) \left[ Q_M(\varepsilon, r) - Q_M(\varepsilon, \infty)\right],\label{eq:Z2M}
\end{equation} 
where we determine $f_\mathrm{ex}(r) = (4\pi^{3})^{-2}\int\mathrm{d}\boldsymbol{\alpha}\,\Theta(r,\boldsymbol{\alpha})$ numerically \cite{Note1}, and with
\begin{align}
  Q_M(\varepsilon, r)=
    &(Z_\mathrm{rot}^{(1)})^2\frac{(2N)!}{(2N-M)!M!}S^{M}\\\notag
    &\times\exp\left(
    -\int_{-\varepsilon}^{\infty} d\varepsilon' \, [\langle B(\varepsilon,r)\rangle_M -M]
  \right).
\end{align}
This central results implies that the partition sum $Z^{(2)}$ can be calculated using the mean number of bound stickers $\langle B(\varepsilon,r)\rangle_M$ for fixed $M\in[1,\,2N-1]$, as a function of the distance $r$ and the binding energy $\varepsilon$.
We obtain $\langle B(\varepsilon,r)\rangle_M$ using a Monte Carlo  algorithm \cite{Note1}.

\begin{figure}[ht!]
\centering
\includegraphics[width=9cm]{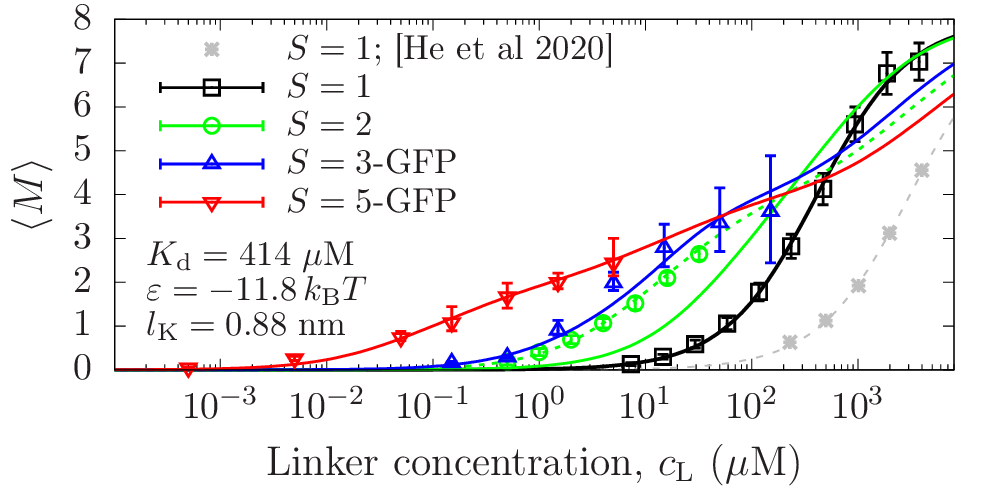}
\caption{The number of bound linkers, $\langle M \rangle$,  against their concentration, $c_\mathrm{L}$, for various numbers of stickers, $S$. The symbols are measured using SPR and Slimfield. The solid curves correspond to the best fit to the $S=1,3,5$ data, and the dashed green curve is the best fit to the $S=2$ data. %The symbols represent the measured data; the solid curves are the theory for $\varepsilon=-11.6\kT$; $l_\mathrm{K}=0.89$ nm; $K_\mathrm$; the dashed curves are best fits
%sequences with $1$ and $2$ (SPR), and with $3$-GFP and $5$-GFP (Slimfield) stickers.
}
\label{fig:titration}
\end{figure}

\paragraph*{Titration of Linkers to Single-Rubisco. --} 
To ex\-pe\-ri\-men\-tal\-ly test the theory, we will pa\-ra\-metrise the model using the concentration-dependent number of bound molecules $\langle M\rangle$ in Eq.~(\ref{eq:Mmean}) for various sequences, and compare predictions on condensation against microscopy observations. 
For these experiments, Rubisco was purified from \emph{C. reinhardtii} and EPYC1 variants with differing sticker numbers ($S =1,\,2,\,3,\,4,\,5$, and Green Fluorescent Protein (GFP) tagged  $3$-GFP,  $5$-GFP) were produced and purified from \emph{E. coli} \cite{Note1}. 
The $S = 1$ and $S = 2$ variants were used as analytes in SPR experiments in a buffer of $50$ mM Tris-HCl and $50$ mM NaCl at pH $8$  \cite{Note1}, in which Rubisco was immobilised on the chip surface and the binding response was determined across titration curves for each variant (Fig.~{~\ref{fig:titration}}).
Variants of EPYC1 containing more than two stickers ($S > 2$) give rise to spontaneous phase separation of Rubisco at concentrations exceeding the critical concentration \cite{Note1} and therefore could not be used in SPR experiments due to their reliance on equilibrium binding.

\begin{figure}[ht!]
\centering
\includegraphics*[width=8cm]{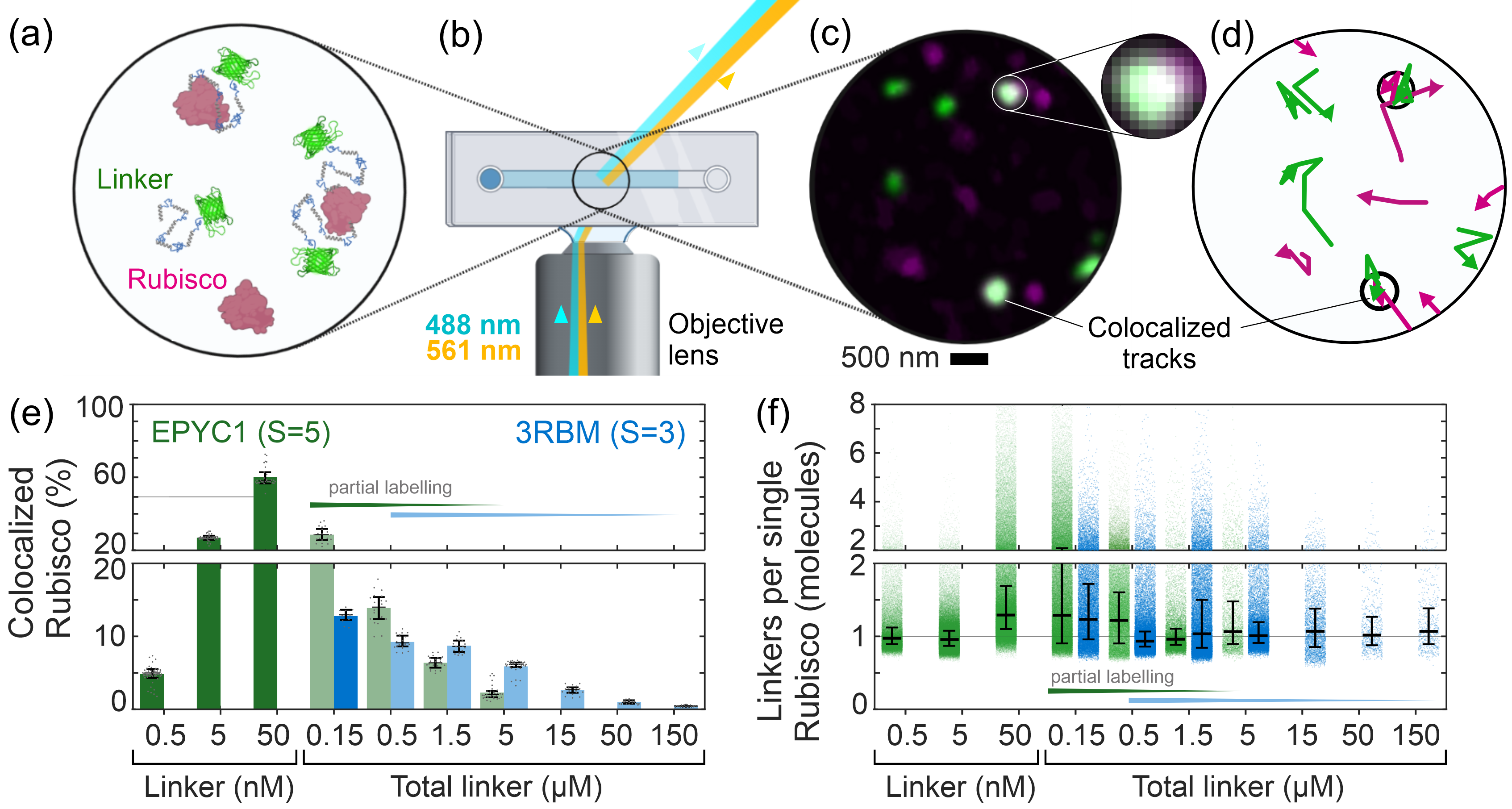}

\caption{Quantitative binding of linker and Rubisco using Slimfield. 
a,b) Rubisco-Atto594 is equilibrated with linker at mutual concentrations insufficient for phase separation, and introduced to a simple microscope chamber. c) Slimfield reveals how assemblies of Rubisco (magenta, max projection) and/or linker-GFP (green) adsorb transiently and non-specifically to the coverglass. d) Rapid molecular motion is reconstructed into tracks and unique colocalisations. e) The fraction of individual Rubiscos with colocalised linker (means: EPYC1-GFP, green; 3RBM-GFP, blue; medians/IQRs: black), $\theta$, increases with visible linker. Partial labelling at total linker $\geq 150$ nM masks the underlying binding curves. f) Non-zero stoichiometries $M_{[>0]}$ of linker-GFP at each Rubisco also rise with labelled linker (EPYC1-GFP, green; 3RBM-GFP, blue; medians/IQRs, boxes). The product of the two results is corrected for partial labelling to yield $\langle M\rangle$, which increases monotonically with total linker (Fig 3).}
\label{fig:Slimfield}
\end{figure}

Before we discuss the curve-fits to the data, we now first  focus on the  titration curves for the $S=3$ and $S=5$ variants that we have measured using Slimfield microscopy \cite{Note1}.  Slimfield is a fluorescence microscopy technique that tracks protein assemblies at millisecond timescales in multiple colours and counts them with single-molecule sensitivity \cite{Plank09}.  Coupled to bespoke tracking analysis \cite{Miller15}, this technique examines and quantifies molecular dynamics \emph{in vitro} \cite{Shepherd21} and \emph{in vivo} \cite{Payne-Dwyer22}. We use this pipeline to identify and co-track individual complexes of labelled Rubisco and/or linker near a coverslip surface without specific binding, at nanomolar concentrations. (Fig.~\ref{fig:Slimfield}a-c).   For these experiments we used the $S=3,\,3$-GFP$,\,5$ and $5$-GFP EPYC1 variants, as well as  Rubisco that was non-specifically labelled with a fluorescent Atto594 dye.   Here, our estimate of $\langle M\rangle$ follows from the expression $\langle M\rangle = \theta 
\langle M_{[>0]}\rangle /\phi_\mathrm{GFP}$, comprising two observable factors: $\theta$, the fraction of detected single Rubisco foci that are colocalised to linker foci, and $\langle M_{[>0]}\rangle$, the average apparent stoichiometry of those colocalised linker foci, then corrected for the visible molar fraction, $\phi_\mathrm{GFP}$, of linker-GFP in total linker.

Low concentrations of Rubisco-Atto594 were used to ensure a dilute spatial distribution of isolated Rubisco foci in the field of view, and mixed with excess linker at a range of total concentrations  ($0.5$ nM $-$ $50$ nM linker at $5$ nM Rubisco, and $150$ nM $-$ $150$ $\mu$M linker at $50$ nM Rubisco). All experiments used the same buffers  as in the SPR experiments.  To maintain identifiable and distinct linker foci, the maximum linker-GFP concentration was fixed at 50 nM ($S=5$-GFP) or 150 nM ($S=3$-GFP), such that higher concentrations were diluted with the corresponding unlabelled linker ($10^{-3}<\phi_\mathrm{GFP}<1$).   In each condition, $>60,000$ tracks each corresponding to a single molecule of Rubisco-Atto594 were detected from $>10$ independent acquisitions (Fig.~\ref{fig:Slimfield}d). 

For the native linker ($S=5$) the proportion of colocalised Rubisco, $\theta$, rises above 50\% with linker concentration (Fig.~\ref{fig:Slimfield}e) indicating partial binding saturation. The concentration at which half of the Rubisco proteins are bound by at least one linker-GFP lies between $5-50$ nM (Fig.~\ref{fig:Slimfield}j, green data), which resembles the binding affinity of $29 \pm 12$ nM   estimated using Fluorescence Correlation Spectroscopy\cite{He23} \cite{Note1}. 
At low labelling fractions $\phi_\mathrm{GFP}$, mostly isolated linker-GFPs are observed at each Rubisco so that the binding response is largely encoded in $\theta/\phi_\mathrm{GFP}$. The binding affinity is weakened for $S=3$, and shifts this characteristic concentration to $\approx 1 \mu$M.

We have curve-fitted  Eq.~(\ref{eq:Mmean}) to all titration curves obtained using SPR and Slimfield data in Fig.~\ref{fig:titration}. For $S=1$, Eq.~(\ref{eq:Mmean}) reduces to $\langle M\rangle = N{c_\mathrm{L}}/({K_\mathrm{d}+c_\mathrm{L}})$, with $K_\mathrm{d}\propto\exp(\varepsilon/\kT)$ the dissociation constant and $c_\mathrm{L}=K_\mathrm{d}\exp((-\varepsilon+\mu)/\kT)$ the concentration of unbound linkers (for low Rubisco concentrations this approximately equals the total concentration $c_\mathrm{L}\approx c_{\mathrm{L},0}$). The curve-fit to our SPR data of the 60-residue $S=1$ fragment yielded $K_\mathrm{d}=414\pm 52$ $\mu$M, and is smaller than the $K_\mathrm{d}\approx 3$ mM  of a 24-residue variant \cite{He20}, albeit under different buffer conditions. 
By simultaneously fitting the model to our $S=3$-GFP and $S=5$-GFP data we found $l_\mathrm{K} = 0.88\pm 0.12$ nm and $\varepsilon = -11.8\pm 0.8\kT$, which we have used to calculate all solid curves in Fig.~\ref{fig:titration}.
The variances are correlated through $\varepsilon\approx 11l_\mathrm{K}^{-0.45}$, and will be used in Fig.~\ref{fig:StickerDependence} to calculate confidence intervals on our predictions for dimerisation concentrations. 
The $S=2$ data displayed a higher binding affinity than expected, and was (non-uniquely) fitted using  $\varepsilon\approx -12.7 l_\mathrm{K}^{-0.55} \kT$ (green dashed curve). 
It is inconclusive if this discrepancy is due to experimental factors (e.g., crosslinking of Rubisco at the surface; influence of fluorescent tags or coverslip, etc.), or if it may point at missing pieces of physics.

\paragraph*{Condensation Microscopy. --} 
To calculate dimerisation diagrams\cite{Note1} for linkers with $S = 2,\, 3,\, 4,\, 5$ stickers we have  used $K_\mathrm{d}=414$ $\mu$M; $\varepsilon=-11.8\kT$, and $l_\mathrm{K}=0.88$ nm  for the best fit to the single-molecule data in Fig.~\ref{fig:titration}, and we have propagated the errors by using $(-12.5\kT$, $0.75$ nm) and $(-11.0\kT$, $1.0$ nm), as informed by the above-discussed relationship $\varepsilon=-11l_\mathrm{K}^{-0.45} \kT$.
Fig.~\ref{fig:StickerDependence}  suggests that the characteristic concentration determined using Eq.(\ref{eq:DimerisationConcentration}) $c_{\mathrm{R},0}+c_{\mathrm{L},0}$, decreases  with a decreasing Kuhn length in agreement with recent claims in a simulation study \cite{GrandPre23}. However, a closer inspection reveals  the $S=2$ molecule is an exception to that, and our full analysis indeed confirms non universality \cite{Note1}.

To experimentally approximate the actual critical concentration for all untagged variants, we have performed condensation assays with a linker fraction fixed to $c_{\mathrm{L},0}/(c_{\mathrm{L},0}+c_{\mathrm{R},0})=0.88,\,0.79,\,0.68,\,0.51$ for $S = 2,\, 3,\, 4,\, 5$, respectively, while the overall concentration $c_{\mathrm{L},0}+c_{\mathrm{R},0}$ was titrated until condensation was observed using microscopy \cite{Note1}. 
We compare the theoretical and the experimental values in Fig.~\ref{fig:StickerDependence}.
 
We find striking agreement between the theory and experiments for the $S=3$ and the $S=4$ variants. However, for the $S=2$ variant, which also showed distinct behaviour in Fig.~\ref{fig:titration}, we did not observe the formation of droplets.
Perhaps surprisingly, the theory predicts an increasing dimerisation concentration for EPYC1 (the $S=5$ variant) compared to the $S=4$ variant. Instead, the measured critical concentration is an order of magnitude lower than predicted. While there is no intuitive explanation for the non-monotonous `magic number' prediction, we speculate the model predictions may be  affected by (1) the intersite distances on Rubisco \cite{GrandPre23}, as well as by the (perhaps too) idealized force-extension model in  Eq.~\ref{eq:StretchPotential}; (2) non-specific interactions that we ignored, but which are needed to explain phenomena such as gelation \cite{Harmon17}; (3) cooperativity (or nucleation) effects: it is not unthinkable that the $S=5$ variant binds more easily to multiple holoenzymes than the shorter variants. 
We anticipate our dimerisation diagrams may inform concentration regimes of interest in large-scale simulations to address these open questions.

\paragraph*{Conclusions. --}  
We have crucially tested `sticker-and-spacer theory'  by quantitatively comparing it to self-assembly properties both in the dilute and concentrated phase. 
The fits of the model to dilute-phase titration curves not only supports the theory, but also enables the measurement of both the sticker binding energy and the Kuhn length of the spacers. These allow for the prediction of dimerisation diagrams, as well as a (crude) estimate for the critical point for condensation. 
By applying this approach to pyrenoids, we have found striking agreements for some linker variants, but also qualitative disagreements that point at open questions in the field. To arrive at these findings, we have developed semi-analytical equations, numerical algorithms, and co-localisation analyses in single-molecule microscopy, see Supplementary Material. We hope this pipeline to be of interest to the wider research on multi-component sticker-spacer systems in soft matter science and the physics of life.

%\begin{acknowledgments} % RevTeX  	
\textit{In memory of Tom McLeish, who was involved in the conceptualisation of the `York Physics of Pyrenoid Project (YP3)' that this research is part of.
The YP3 consortium and External Advisory Board are thanked for fruitful discussions. Johan Lukkien is thanked for consultation on kMC algorithms and software architectures. Philipp Girr is  thanked for helping to purify/label proteins. The University of York Bioscience Technology Facility is thanked for SPR and microscopy support. 
Supported  by EPSRC (EP/W024063/1); UKRI FLF (MR/T020679/1), BBSRC-NSF/BIO grant (BB/S015337/1) and Carbon Technology Research Foundation grant (AP23-1\_023) to LCMM; 
and a BBSRC DTP2 (BB/M011151/1a) to JB, LCMM, ML.}
 %\end{acknowledgments}   % RevTeX

\bibliography{ms} 
%\printbibliography
%\end{widetext}
%\lipsum[1-3]
\end{document}